\documentclass[a4paper]{jpconf}
\usepackage{graphicx}
\usepackage{wrapfig}
\begin{document}
\newcommand{\pip}          {$\pi^{+}$}
\newcommand{\pim}          {$\pi^{-}$}
\newcommand{\kap}          {K$^{+}$}
\newcommand{\kam}          {K$^{-}$}
\newcommand{\pbar}         {$\rm\overline{p}$}
\newcommand{\degree}       {$^{\rm o}$}
\newcommand{\dg}           {\mbox{$^\circ$}}
\newcommand{\dedx}         {d$E$/d$x$}
\newcommand{\dndy}         {d$N$/d$y$}
\newcommand{\pp}           {pp}
\newcommand{\ppbar}        {\mbox{$\mathrm {p\overline{p}}$}}
\newcommand{\PbPb}         {\mbox{Pb--Pb}}
\newcommand{\pPb}          {\mbox{p--Pb}}
\newcommand{\AuAu}         {\mbox{Au--Au}}
\newcommand{\pA}           {\mbox{p--A}}
\newcommand{\dA}           {\mbox{d--A}}
\newcommand{\AAA}          {\mbox{A--A}}
\newcommand{\pseudorap}    {\mbox{$\left | \eta \right | $}}
\newcommand{\Nch}          {\ensuremath{N_\mathrm{ch}}}
\newcommand{\dNdeta}       {\ensuremath{\mathrm{d}N_\mathrm{ch}/\mathrm{d}\eta}}
\newcommand{\dNdpt}        {\ensuremath{\mathrm{d}N_\mathrm{ch}/\mathrm{d}\ensuremath{p_{\rm t}}}}
\newcommand{\dNdetatr}     {\mathrm{d}N_\mathrm{tracklets}/\mathrm{d}\eta}
\newcommand{\dNdetar}[1]   {\mathrm{d}N_\mathrm{ch}/\mathrm{d}\eta\left.\right|_{|\eta|<#1}}
\newcommand{\lum}          {\, \mbox{${\rm cm}^{-2} {\rm s}^{-1}$}}
\newcommand{\barn}         {\, \mbox{${\rm barn}$}}
\newcommand{\m}            {\, \mbox{${\rm m}$}}
\newcommand{\s}            {\ensuremath{\sqrt{s}}}
\newcommand{\jpsi}           {\ensuremath{J/\psi}}
\newcommand{\psip}           {\ensuremath{\psi (2S) }}
\newcommand{\ups}[1]           {\ensuremath{\Upsilon (#1S) }}
\newcommand{\rppb}          {\ensuremath{R_{\rm pPb} }}
\newcommand{\rdau}          {\ensuremath{R_{\rm dAu}}}
\newcommand{\mt}           {\ensuremath{m_{\rm t}}}
\newcommand{\snn}          {\ensuremath{\sqrt{s_{\rm NN}}}}
\newcommand{\snnbf}        {\ensuremath{\mathbf{{\sqrt{s_{\mathbf NN}}}}}}
\newcommand{\sonly}        {\ensuremath{\sqrt{s}}}
\newcommand{\Nanc}         {\ensuremath{N_\mathrm{ancestors}}}
\newcommand{\TAB}          {\ensuremath{T_\mathrm{AA}}}
\newcommand{\Nspec}        {\ensuremath{N_\mathrm{spec}}}
\newcommand{\Npart}        {\ensuremath{N_\mathrm{part}}}
\newcommand{\avNpart}      {\ensuremath{\langle N_\mathrm{part} \rangle}}
\newcommand{\avNpartdata}  {\ensuremath{\langle N_\mathrm{part}^{\rm data} \rangle}}
\newcommand{\avNpartgeo}   {\ensuremath{\langle N_\mathrm{part}^{\rm geo} \rangle}}
\newcommand{\Ncoll}        {\ensuremath{N_\mathrm{coll}}}
\newcommand{\avNcoll}      {\ensuremath{\langle N_\mathrm{coll} \rangle}}
\newcommand{\avNcolldata}  {\ensuremath{\langle N_\mathrm{coll}^{\rm data} \rangle}}
\newcommand{\avTAB}        {\ensuremath{\langle T_\mathrm{AA} \rangle}}
\newcommand{\Dnpart}       {\ensuremath{D\left(\Npart\right)}}
\newcommand{\DnpartExp}    {\ensuremath{D_{\rm exp}\left(\Npart\right)}}
\newcommand{\dNdetapt}     {\ensuremath{\dNdeta\,/\left(0.5\Npart\right)}}
\newcommand{\dNdetaptr}[1] {\ensuremath{\dNdetar{#1}\,/\left(0.5\Npart\right)}}
\newcommand{\dNdetape}     {\left(\ensuremath{\dNdeta\right)/\left(\avNpart/2\right)}}
\newcommand{\dNdetaper}[1] {\ensuremath{\dNdetar{#1}\,/\left(\avNpart/2\right)}}
\newcommand{\etac}         {\ensuremath{\eta_{\rm cms}}}
\newcommand{\dNdetac}      {\ensuremath{d}N_\mathrm{ch}/\ensuremath{d}\etac}
\newcommand{\dndydpt}      {\ensuremath{{\rm d}^2N/({\rm d}y {\rm d}p_{\rm t})}}
\newcommand{\abs}[1]       {\ensuremath{\left|#1\right|}}
\newcommand{\signn}        {\ensuremath{\sigma^{\rm inel}_{\rm NN}}}
\newcommand{\vz}           {\ensuremath{V_{z}}}
\newcommand{\stat}         {({\it stat.})}
\newcommand{\syst}         {({\it sys.})}
\newcommand{\Fig}[1]       {Fig.~\ref{#1}}
\newcommand{\Ref}[1]       {Ref.~\cite{#1}}
\newcommand{\green}[1]     {\textcolor{green}{#1}}
\newcommand{\blue}[1]      {\textcolor{blue}{#1}}
\newcommand{\red}[1]       {\textcolor{red}{#1}}
\newcommand{\white}[1]     {\textcolor{white}{#1}}
\newcommand{\pT} {\ensuremath{p_{\rm T}}}
\newcommand{\pTa} {\ensuremath{p_{\rm T, assoc}}}
\newcommand{\pTt} {\ensuremath{p_{\rm T, trig}}}
\newcommand{\ncoll} {\ensuremath{N_{\rm coll}}}
\newcommand{\nhard} {\ensuremath{n_{\rm hard}}}
\newcommand{\meanpt} {\ensuremath{\langle \pT \rangle}}
\title{Multiple Parton Interactions with ALICE: \\
from pp to p--Pb}

\author{Andreas Morsch \\ for the ALICE Collaboration}

\address{CERN, 1211 Geneva, Switzerland}

\ead{andreas.morsch@cern.ch}

\begin{abstract}
The study of multiplicity dependent di-hadron angular correlations 
allows us to assess the contribution of multiple-parton interactions 
to particle production. We will review these measurements in pp and p--Pb 
collisions with the ALICE detector at the LHC and discuss the results in 
the context of centrality determination and other multiplicity dependent 
observables in p--Pb. 

\end{abstract}

\section{Introduction}
The concept of multiple parton--parton interactions (MPI) provides the theoretical basis to understand 
global event properties of non-diffractive, minimum bias pp collisions. 
The model also allows for a straightforward interpretation of the fact that
at high $\sqrt{s}$ the leading order
cross-section for $2 \rightarrow 2$ parton scatterings with momentum transfer $Q > Q_{\rm min} \gg \Lambda_{\rm QCD}$ 
exceeds the total pp cross-section at a range of $Q_{\rm min }$-values where perturbative QCD is applicable \cite{mpi_sjostrand}. 
At LHC energies, this happens already at $Q_{\rm min} \approx 4$~GeV/$c$
\cite{Sjostrand:2004pf, Sjostrand:2004ef, Bahr:2008wk}.
In a naive 
factorisation approach the mean number of scatterings per event (\nhard)
is equal to the ratio of the hard and the total cross-section, $\sigma_{\rm hard}/ \sigma_{\rm tot}$.
In more realistic phenomenological models like PYTHIA6 \cite{Sjostrand:2006za},  PYTHIA8 
\cite{Sjostrand:2007gs} and HERWIG++ \cite{Bahr:2008wk}, the colour screening effect regularises  the divergence ($\propto 1 / \pT^4$) of the hard cross-section
at low \pT\ ($< 2$~GeV/$c$).  Moreover, being limited by energy and momentum conservation
\nhard\ can not reach arbitrarily large values. 
In particular, these models implement a pp impact parameter dependence which 
explains the so called jet-pedestal effect in the underlying event of hard
collisions \cite{impact}.

\begin{figure}[htbp]
\begin{minipage}{0.48\linewidth}
\centering
\includegraphics[scale=0.4]{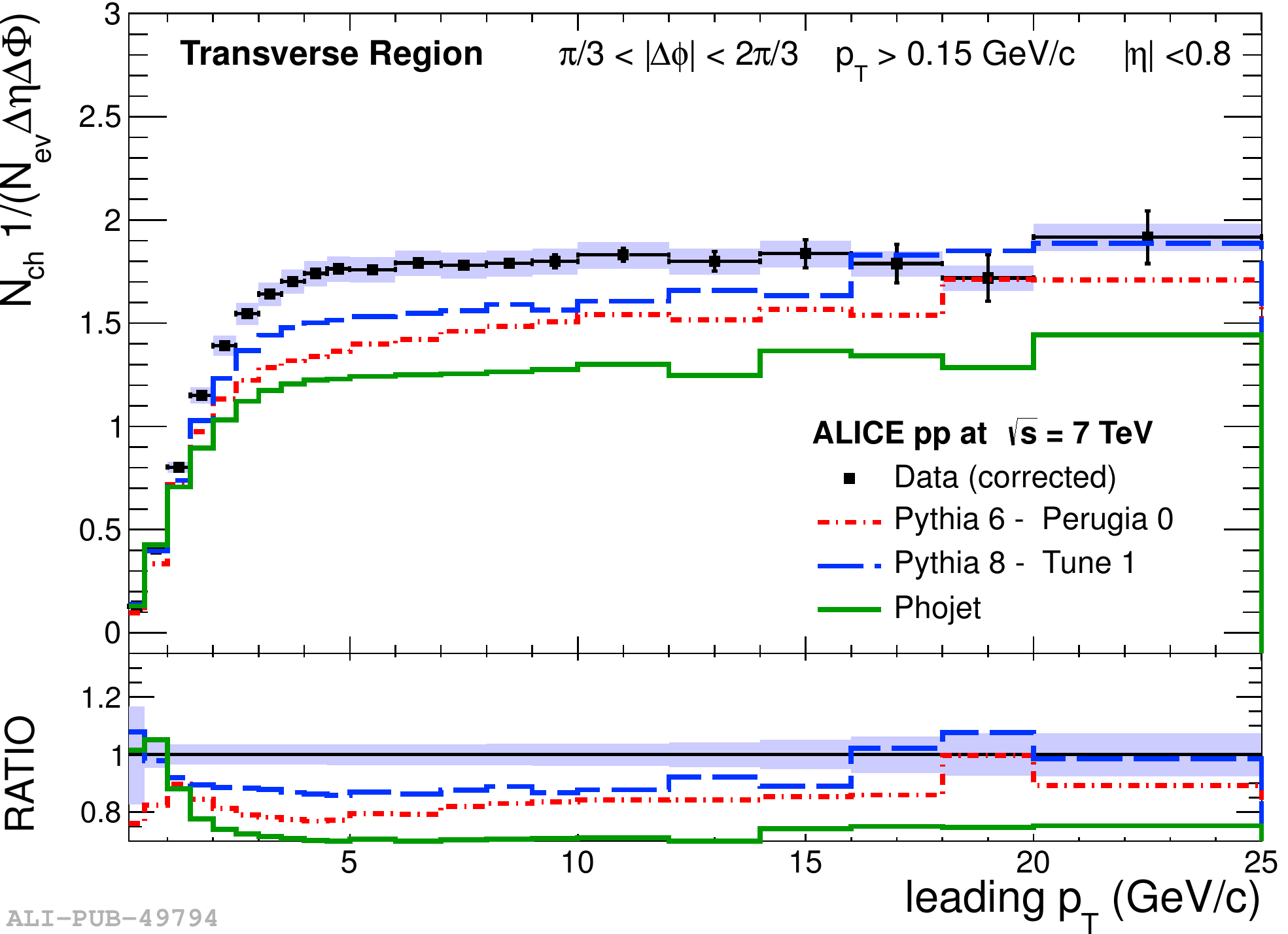}
\caption{Number density in the transverse region 
of leading particles as the function of the leading particle $\pT$
for pp collisions  at $\sqrt{s} = 7$~TeV \cite{ALICE:2011ac}.}
\label{fig:ue}
\end{minipage}
\hspace{0.01\linewidth}
\begin{minipage}{0.48\linewidth}
\centering
\includegraphics[scale=0.4]{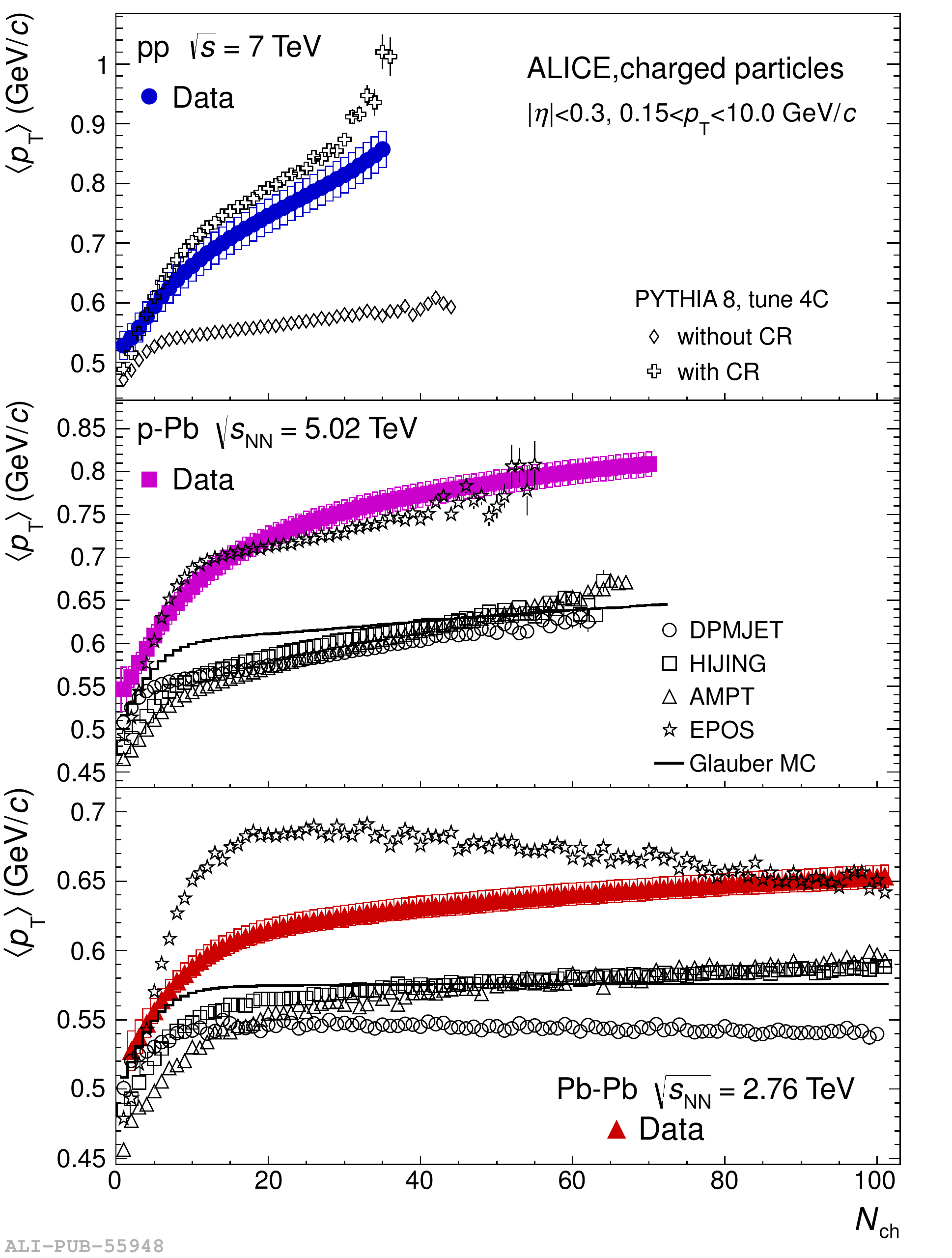}
\caption{The mean particle transverse momentum as a function of multiplicity in $\abs{\eta} < 0.3$ 
for pp collisions at  $\sqrt{s} = 7$~TeV (upper panel), p--Pb collisions at  $\snn = 5.02$~TeV (middle panel) 
and Pb--Pb collisions at $\snn = 2.76$~TeV  (lower panel) \cite{Abelev:2013bla}.}
\label{fig:meanpt}
\end{minipage}
\end{figure}

The underlying event is conventionally defined as the region transverse to the leading jet or particle in 
an event, $\pi/3 < |\Delta \varphi | < 2\pi/3$, where $\Delta \varphi$ is the azimuthal distance to the leading object. 
Figure  \ref{fig:ue} 
shows for pp collisions at $\sqrt{s}=7$~TeV
the measured charged particle density in the transverse region as a function of the leading particle transverse 
momentum ($\pT$) \cite{ALICE:2011ac}.
The particle density exhibits a steep rise at low $\pT$ until 
$\pT \approx 4$~GeV/$c$ where a plateau is reached. In the framework of MPI, events with a higher leading particle 
$\pT$ correspond to more central pp collisions in which the probability for additional uncorrelated hard 
scatterings is enhanced. This effect is also well known in nucleus--nucleus collisions and described by the 
proportionality of the yield from hard processes to the nuclear overlap function.
For example, in Pb--Pb collisions about half of 
yield of particles or jets from hard processes is found in the 10\% most central collisions.
Consequently selecting high-$\pT$ particles one selects also preferentially more central collisions 
until a maximal centrality bias is reached.

Coherence effects between multiple scatterings, called colour re-connections (CR),  
have been introduced in order to explain the steep rise of the particle
mean $\pT$ as a function of the event multiplicity $N_{\rm ch}$ observed in pp collision \cite{Skands:2007zg, Gieseke:2012ft}.
Figure \ref{fig:meanpt} (upper panel) shows the measured $\langle \pT \rangle (N_{\rm ch})$ compared to the results from 
PYTHIA8 simulations with and without CR \cite{Abelev:2013bla}.
The model without CR clearly fails 
to describe the data, whereas by including this effect the agreement is quite good.
The EPOS LHC Monte Carlo generator \cite{Pierog:2013ria} which includes collective hydrodynamic flow for small systems like pp,  is 
also in good agreement with $\langle \pT \rangle (N_{\rm ch})$ and the underlying event measurements at the LHC.

At LHC energies, the large number of initial hard parton--parton scatterings is a common feature 
of high-multiplicity pp, p--Pb and Pb--Pb collisions. 
In the MPI model, high-multiplicity pp collisions arise from low-impact parameter collisions and statistical upward 
fluctuations of the number of MPIs per event. Depending on the position and width of the 
pseudorapidity ($\eta$) window $\Delta \eta$ 
in which multiplicity is counted these
events are also expected to contain harder than average  partonic collisions 
and with partons fragmenting into a larger than
average number of hadrons (fragmentation bias). 
In Pb--Pb, the mean number of initial parton--parton scatterings is determined almost entirely by the 
collision centrality. 
Additional biases are weak. 
Proton-lead collisions can be expected to lie in between these two cases. 
In models that treat p--Pb collisions as independent p--N collisions,
the number of parton--parton scatterings is expected to be determined by the p--Pb and p--N centralities.
Therefore a detailed insight into MPI related effects in pp collisions is needed to understand pp as a reference for p--Pb.

\section{The structure of high multiplicity pp collisions}
\subsection*{Two-Particle Azimuthal Correlations}
Two-Particle azimuthal correlations represent a powerful tool to understand the origin of high-multiplicity 
pp collisions. Such studies involve measuring the distributions of the relative azimuthal angle $\Delta \varphi$
between pairs of particles: a "trigger" particle in a certain transverse momentum $\pTt$-interval and an "associated"
particle in a $\pTa$-interval. 
In these correlations, the fragmentation products of parton-parton scatterings manifest themselves as 
characteristic near-side ($\Delta \varphi = 0$) peak- and away-side ($\Delta \varphi = \pi$) ridge-structures.
The number of correlated particles per trigger particle is defined as the yield of particles in the peaks 
over the uncorrelated background which is constant in $\Delta \varphi$.
In models describing high-multiplicity pp events as superpositions of multiple scatterings 
this normalised  yield is expected to be constant or to increase as a function of multiplicity. 
In the case where a soft component in the \pT -range of the trigger becomes dominant at high multiplicities, 
the yield per trigger-particle decreases.

\begin{figure}[htbp]
\begin{minipage}{0.48\linewidth}
\centering
\includegraphics[scale=0.4]{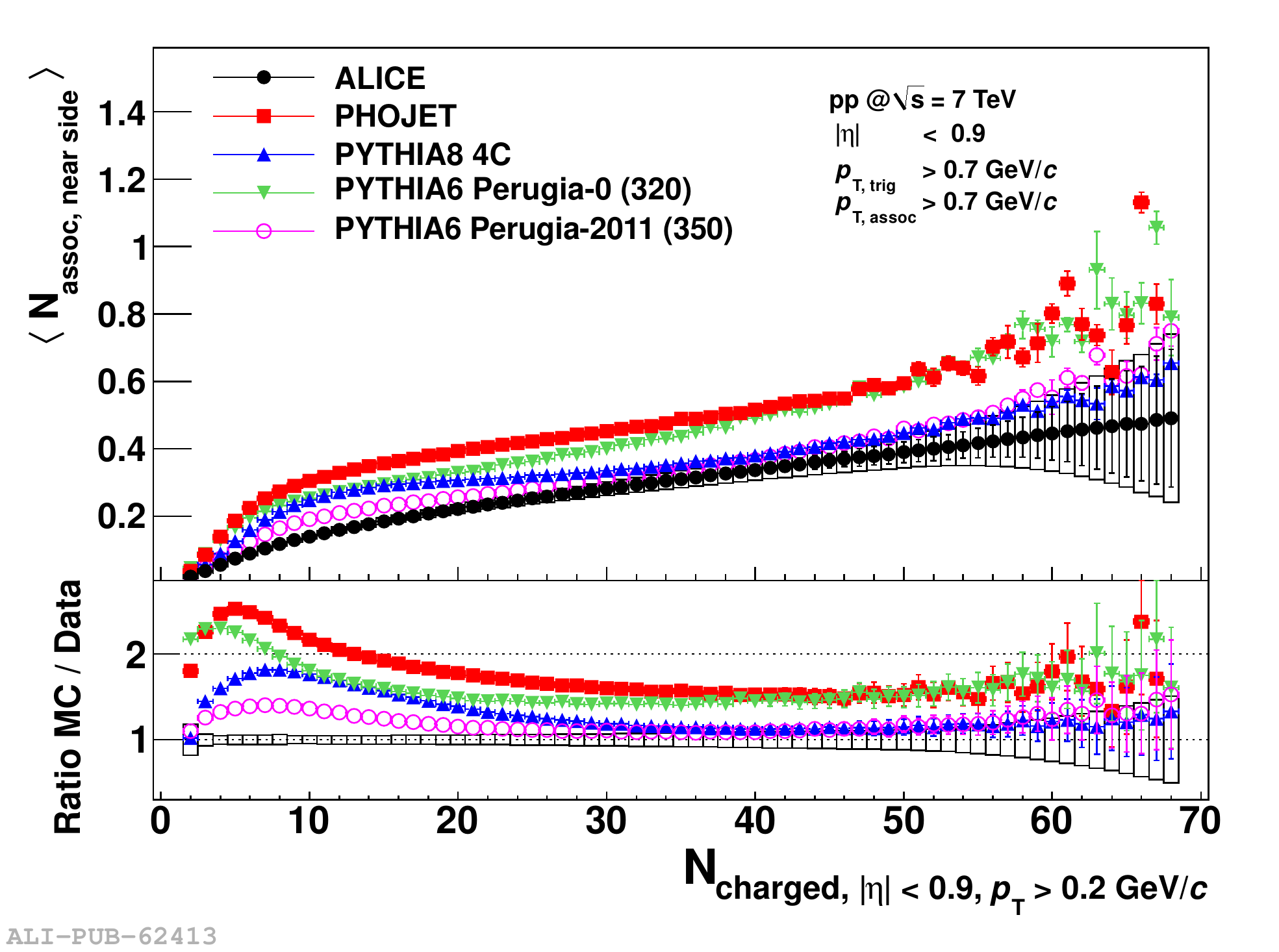}
\caption{ Number of associated particles per trigger particle on the nearside from two-particle azimuthal correlations with 
 $\pTt, \pTa > 0.7$~GeV/$c$ in pp collisions at $\sqrt{s} = 7$~TeV \cite{Abelev:2013sqa}.}
\label{fig:yieldpp}
\end{minipage}
\hspace{0.0\linewidth}
\begin{minipage}{0.48\linewidth}
\centering
\includegraphics[scale=0.4]{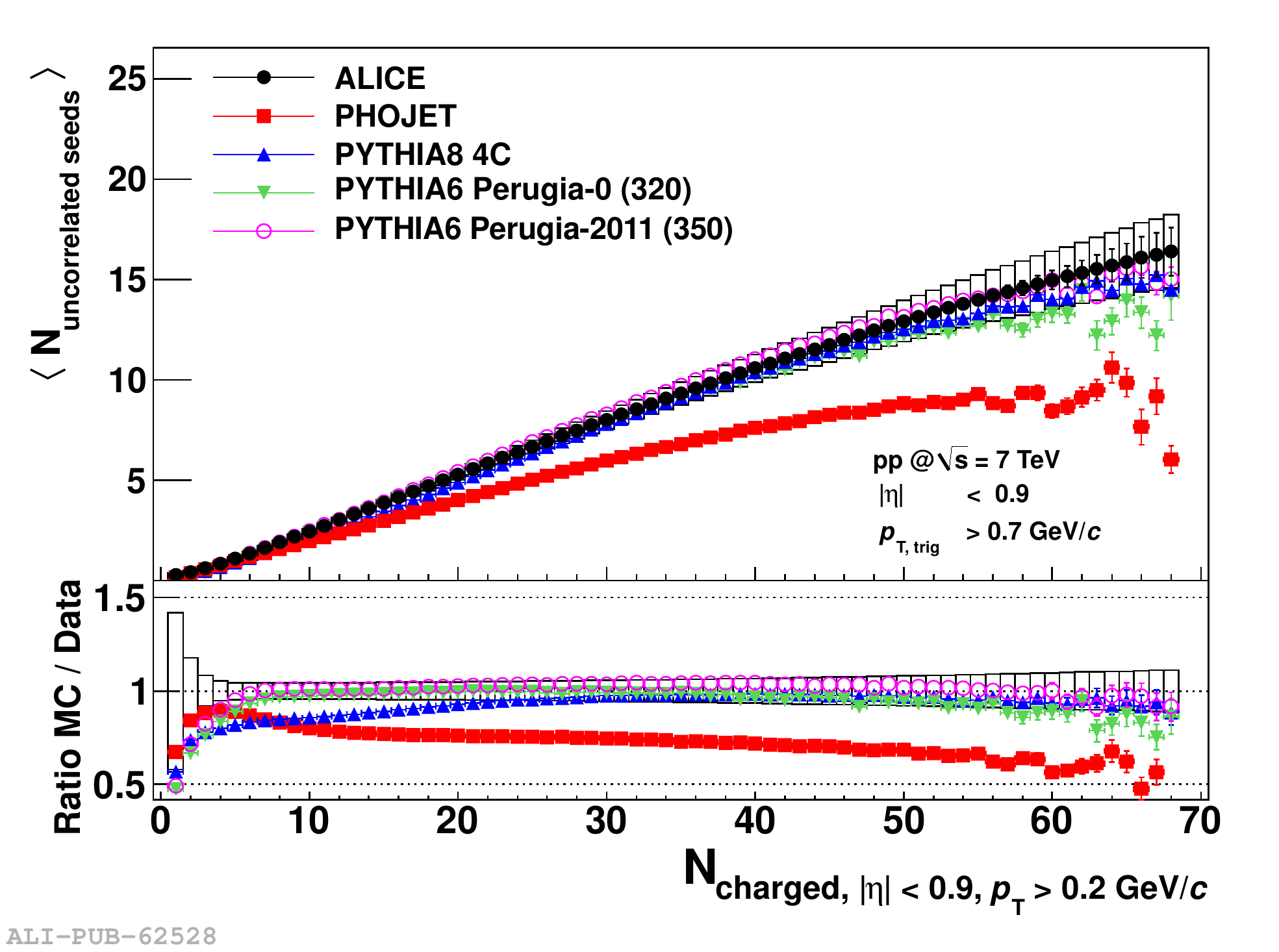}
\caption{
Number of uncorrelated seeds as defined in the text from two-particle azimuthal correlations with 
 $\pTt, \pTa > 0.7$~GeV/$c$ in pp collisions at $\sqrt{s} = 7$~TeV \cite{Abelev:2013sqa}.
} 
\label{fig:uncorseed}
\end{minipage}
\end{figure}

In pp collisions at LHC energies, an increase of the near- and away-side yields as a function of the charged particle multiplicity is observed
(see Fig. \ref{fig:yieldpp}) \cite{Abelev:2013sqa}. This rise is well described by recent PYTHIA6 tunes and PYTHIA8 and in these 
models it is due to the increase of the number of MPIs and the increase of the average $Q^2$ of the collisions.
The number of observed trigger particles depends on the number of initial partons and the number of fragments
per parton. Since the latter rises with multiplicity, a non linear increase of the number of trigger particles
with multiplicity is observed. Hence, the number of trigger particles is not a good measure of the number of MPIs 
contributing to the event multiplicity. 

In order to reduce the sensitivity to parton fragmentation, ALICE studies the
number of uncorrelated seeds ($N_{\rm uncorrelated \, seeds}$), defined as the ratio of the number of trigger particles to 
the sum of near and away-side yields  per trigger particle plus one.   
This ratio increases almost linearly with multiplicity and indicates an onset of a saturation 
at the highest multiplicities (Fig. \ref{fig:uncorseed}).
 In PYTHIA, $N_{\rm uncorrelated \, seeds}$ is proportional to the number of MPIs and the saturation
is related to a steep drop of the probability distribution
for the number of MPIs (for example by 4 orders of magnitude between 12 and 24 for PYTHIA6 tune Perugia-0).

\subsection*{Sphericity}
\begin{figure}[htbp]
\centering
\includegraphics[scale=0.8]{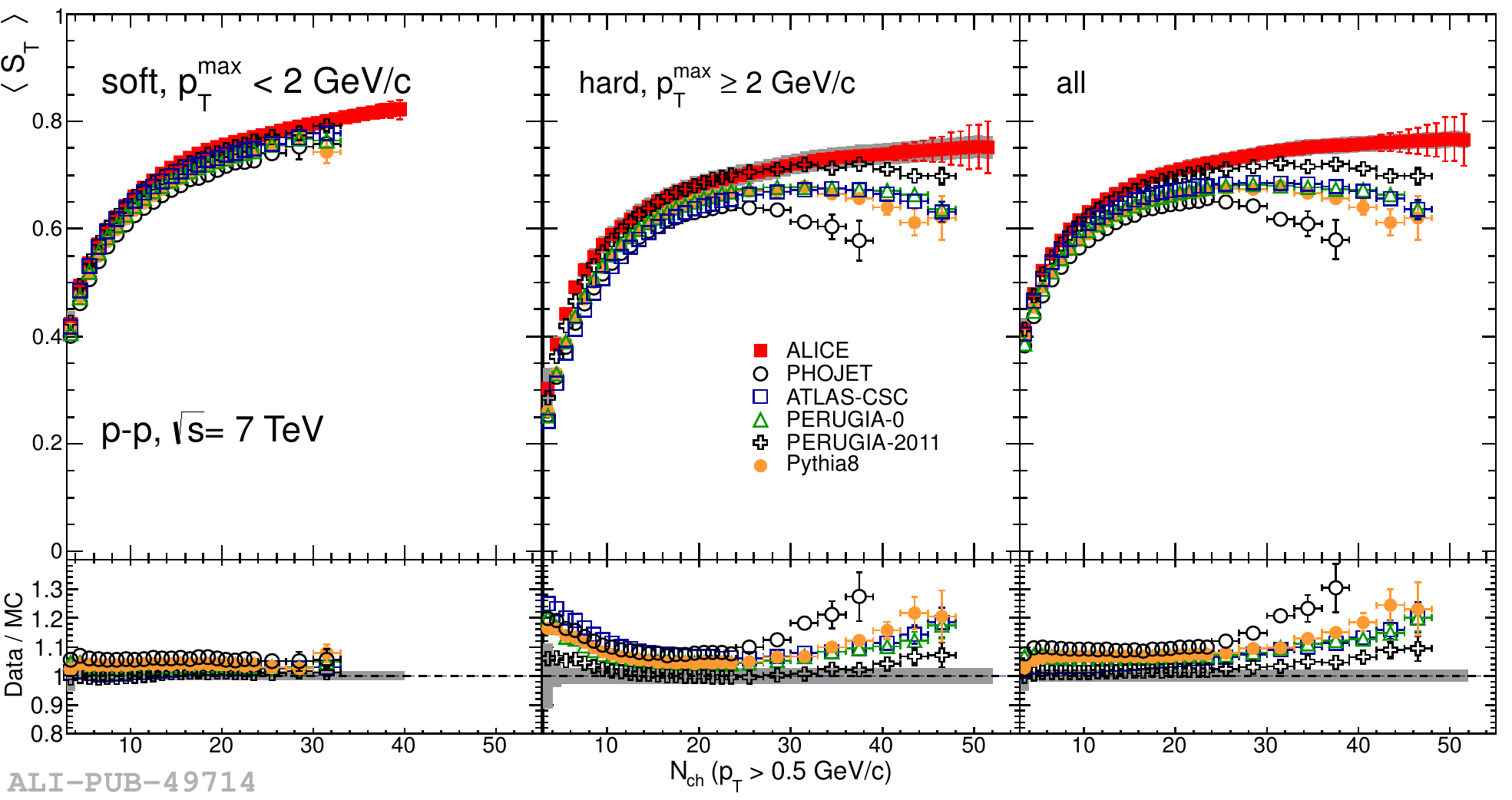}
\caption{The transverse sphericity as a function of multiplicity in pp collisions at $\sqrt{s} = 7$~TeV 
for soft events defined as events not containing any particle with $\pT> 2$~GeV/$c$ (left panel), hard events defined as events containing at least one particle with $\pT> 2$~GeV/$c$ (middle panel) and all events (right panel)
\cite{Abelev:2012sk}.}
\label{fig:sphericity}
\end{figure}

Another way to characterise the structure of high-multiplicity pp events is the transverse
sphericity. This event-by-event observable can vary between 0 (jet like event) and 1 (particle 
distribution isotropic in azimuth).
In data this observable shows a strong rise as a function of multiplicity until it reaches a plateau value 
(Fig. \ref{fig:sphericity}) \cite{Abelev:2012sk}.
The rise is well reproduced by Monte Carlo event generators. However, at the highest multiplicities 
they exhibit a turnover to lower sphericity values 
indicating a stronger (higher $\pT$) jet component as compared to data.

\subsection*{Heavy Flavor yields}
\begin{figure}[htpb]
\begin{minipage}{0.48\linewidth}
\centering  
\includegraphics[scale=0.4]{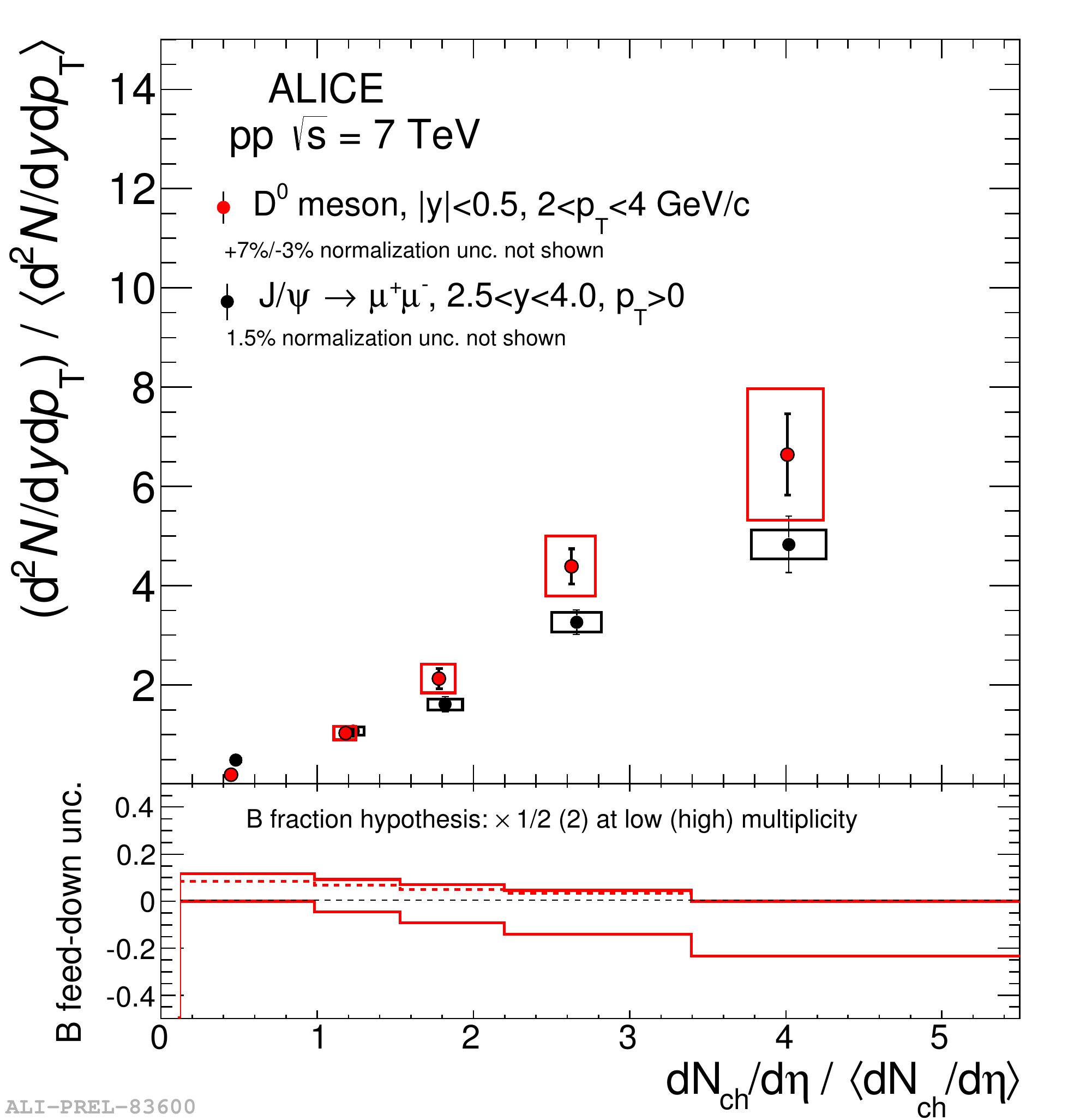}
\end{minipage}
\begin{minipage}{0.48\linewidth}
\caption{${\rm D}^0$-meson self normalised yields vs multiplicity in $2 < \pT < 4$~GeV/$c$ in pp collisions at 7~TeV compared to ${\rm J/psi} \rightarrow  \mu^+  \mu^-$.
Multiplicity was measured via the number of tracklets (segments formed by two hits in different layers of the pixel detector (SPD) compatible with the reconstructed primary vertex)
and is expressed as charged particle density normalised to its mean value.
}
\label{fig:D}
\end{minipage}
\end{figure}

An important test of the MPI origin of high multiplicity events is the study of heavy flavour yields as a function of
multiplicity. Heavy quarks ($c, \, b$)  are created in hard processes with a minimum momentum transfer of $Q > 2m_{\rm Q} \gg \Lambda_{\rm QCD}$
and they are very rarely created during hadronization of a different parton species. Hence, heavy flavour hadrons are ideal tags for hard processes down 
to zero \pT . Under the assumption that the number of MPIs is proportional to the hard cross section and that
the soft particle  multiplicity scales with the number of MPIs one expects that the yield from
 any hard sub-process increases with multiplicity. In ALICE, this behaviour as been verified for D-meson and ${\rm J}/\psi$ 
 production (see Fig. \ref{fig:D}).

\section{MPIs in p--A}
In p--A collisions the number of MPIs is proportional to the number of binary nucleon--nucleon 
collisions (\ncoll ). These MPIs overlap in a transverse collision area similar to the one in pp. However, 
in comparison to pp, much larger number of MPIs can be reached.
It is plausible  that effects related to MPIs observed in pp are also relevant in p--A collisions.
For example effects due to colour reconnections could be enhanced in p--A due to the larger number of MPIs.

Using the same factorisation approach as in pp and assuming that p--A collisions can be described as an independent 
superposition of pp collisions one can write for the mean number of MPIs in p--A:
\begin{equation}
\langle \nhard \rangle _{\rm pA}  = \langle N_{\rm coll} \rangle _{\rm MB}  \langle \nhard \rangle _{\rm pp}
\end{equation}
This implies that in general, particle yields  from hard processes $Y_{\rm hard}$  
normalised by the number of binary collisions scale like
\begin{equation}
{{Y_{\rm hard}} \over { \langle \ncoll^{\rm cent}  \rangle }}  \propto {{ \langle \nhard \rangle _ {\rm pN} }  \over
{\langle \nhard \rangle _{\rm pp}}}.
\end{equation}
For centrality integrated p--A collisions the ratio $\langle \nhard \rangle ({\rm pN}) /
\langle \nhard \rangle  ({\rm pp})$ is unity. However it is important to note that in this framework it can
deviate from unity in centrality event classes. This fact has to be taken into account when studying cold 
nuclear effects comparing p--A results to the the pp baseline. Two examples, the nuclear modification factor and \meanpt\ as a function of multiplicity   are discussed in the following sections.

\subsection*{MPIs and centrality selection in {\rm p--A}}
In general, centrality is defined via estimators that depend monotonically 
on the number of nucleon-nucleon collisions, e.g. multiplicity and summed energy in a 
certain pseudo-rapidity range. 
ALICE employs various sub-detector systems covering disjunct pseudo-rapidity ranges to estimate centrality. 
$CL1$ is the number of clusters counted in the 2nd layer of the Silicon Pixel Detector covering $\abs{\eta} < 1.4$. 
$V0A$ and $V0C$ are the summed amplitudes measured by a pair of scintillator arrays covering 
$2.8 < \eta < 5.1$ and $-3.7 < \eta < -1.7$, respectively. 
Centrality classes are defined as percentiles of the multiplicity/summed-amplitude distributions.

\begin{figure}[htbp]
\begin{minipage}{0.48\linewidth}
\centering
\includegraphics[scale=0.4]{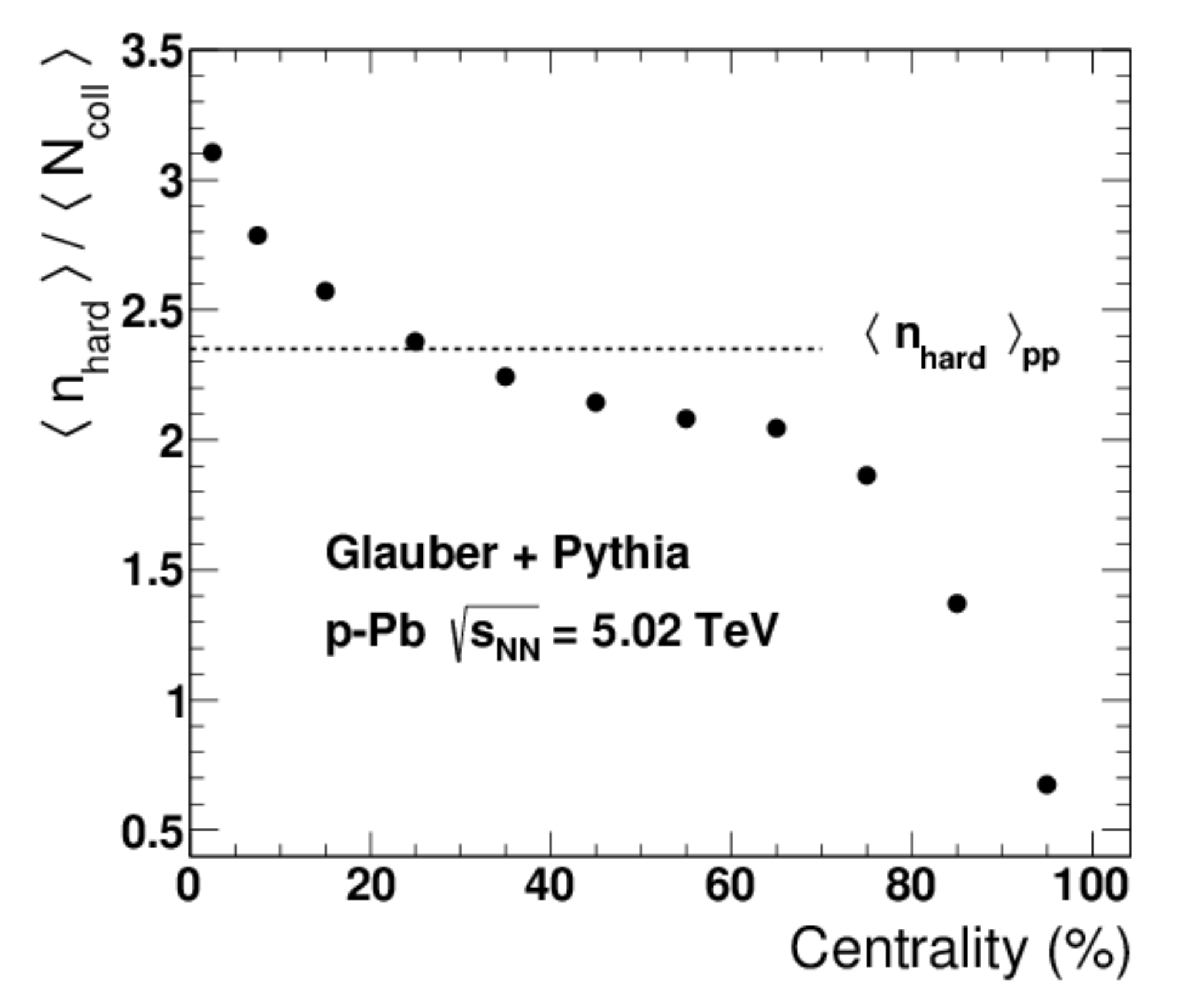}
\caption{Average number of hard collisions per binary p--nucleon collision as a function of centrality from 
the G-PYTHIA calculation described in the text.}
\label{fig:nhard}
\end{minipage}
\hspace{0.0\linewidth}
\begin{minipage}{0.48\linewidth}
\centering
\includegraphics[scale=0.4]{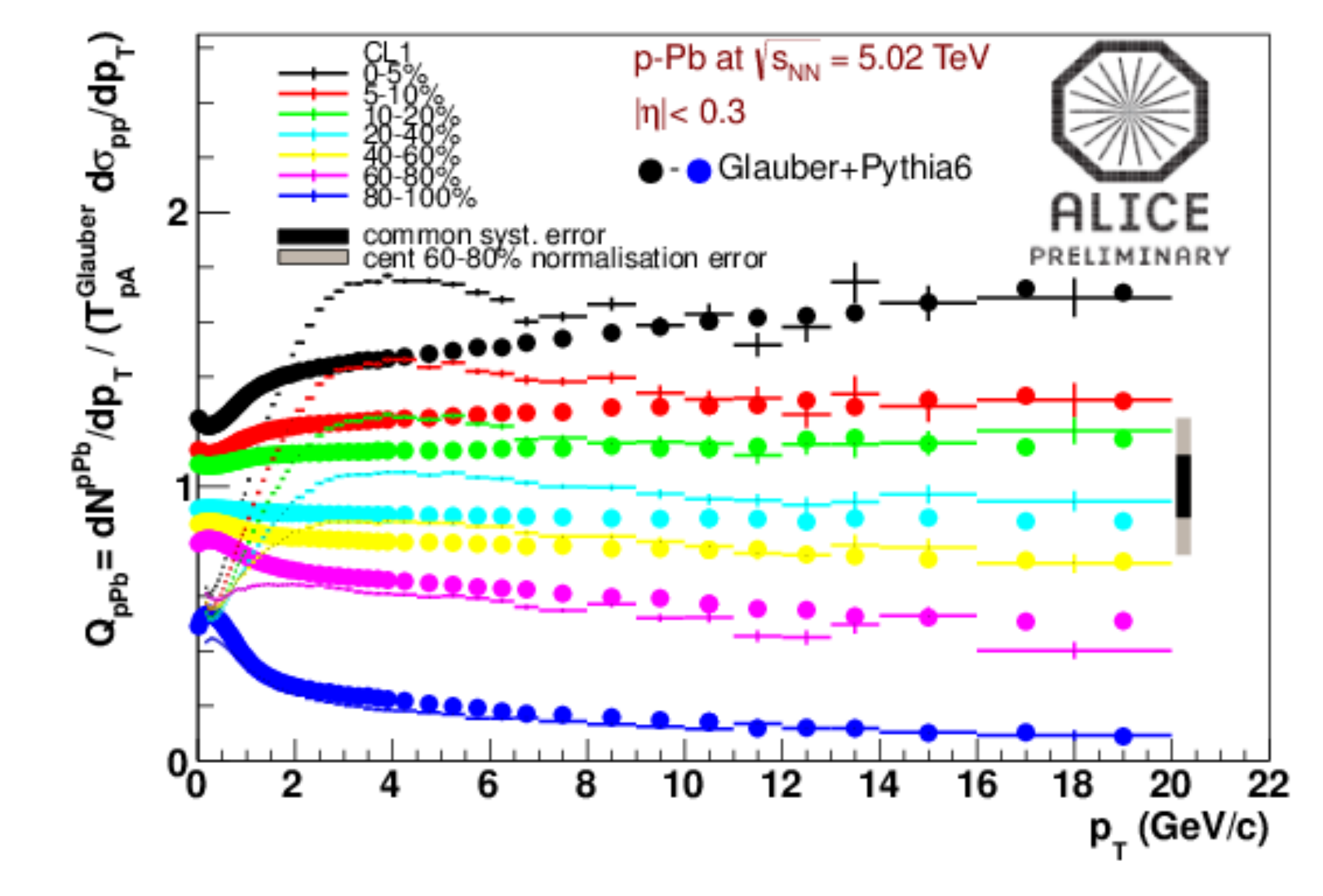}
\caption{$Q_{\rm pPb}$ for the CL1 centrality estimator compared to the G-PYTHIA 
calculation as described in the text for p--Pb collisions at $\snn = 5.02$~TeV .}
\label{fig:qppb}
\end{minipage}
\end{figure}

In order to study the influence of the centrality selection on MPIs in a coherent superposition of p--N collisions
we coupled the PYTHIA6 event generator to a p--Pb Glauber MC calculation (G-PYTHIA). 
For each MC Glauber event PYTHIA is used
\ncoll\ times to generate \ncoll\ independent pp collisions. The resulting charged particle multiplicity distribution 
in the above mentioned $\eta$ ranges are used to define centrality classes similar to the ones used in data \cite{fortheALICE:2013xra}.
Figure \ref{fig:nhard} shows the average number of hard scatterings ($\langle n_{\rm hard} \rangle$) per binary collision as a function of
centrality. In the 20\% most central collisions  $\langle n_{\rm hard} \rangle$ per \ncoll\ is higher than the centrality averaged 
value and in the most peripheral collisions this ratio is far below this average. 
As discussed above this bias has  consequences for the binary scaling of hard processes. 
Figure \ref{fig:qppb} shows the nuclear modification factor $Q_{\rm pPb}^{\rm CL1}$ which is the ratio of the inclusive
charged particle spectrum measured in p--Pb to the pp reference scaled by \ncoll . ALICE does not use the
conventional notation $R_{\rm pPb}$ to indicate the biased nature of the observable. 
At high \pT\  ($>6$~GeV/$c$), the centrality dependence of $Q_{\rm pPb}^{\rm CL1}$ reflects the bias on $\langle n_{\rm hard} \rangle$. In addition, in central (peripheral) collisions, $Q_{\rm pPb}^{\rm CL1}$ rises (decreases)
with \pT . This is the consequence of an additional autocorrelation bias. Hard processes contribute themselves to
the multiplicity. This contribution increases with the hardness of the scattering. Hence pp events containing a hard scattering are classified more central than an average collision.
At high \pT\ the G-PYTHIA model of independent collisions describes the data well. 
Global particle production scales with the number of participants and as we will see later in the  low- and intermediate \pT\ region effects reminiscent of collective behaviour in A--A are observed.
Hence, as expected our simple model does not describe the low- and intermediate \pT\ region. One important observation in this region that will be discussed in the next section is the increase of the average transverse momentum \meanpt\ with multiplicity.


\subsection*{Multiplicity dependent measurements in {\rm p--Pb}}
Figure \ref{fig:meanpt} shows the average transverse momentum as a function of multiplicity for pp collisions at
$\sqrt{s} = 7$~TeV, p--Pb at $\snn = 5.02$~TeV and Pb--Pb at $\snn = 2.76$~TeV \cite{Abelev:2013bla}.
From measurements  of, $\meanpt (N_{\rm ch})$ in pp at several energies we expect that the $\sqrt{s}$-dependence 
of this observable is weak and, hence, we assume the results for the three collision systems can be directly
compared. 
With respect to Pb--Pb, in p--Pb, \meanpt\ shows a much stronger 
increase with multiplicity following the pp data up to $N_{\rm ch} = 14$. 
Note that multiplicities around 14
correspond to typical p--Pb collision, whereas  
pp collisions at this multiplicity are already strongly biased ($N_{\rm ch} > 14$
corresponds to 50\% (10\%) of the p--Pb (pp) cross-section).
As discussed in the introduction, in pp this rise cannot be explained by an independent superposition of multiple parton scatterings
and can be attributed to the effect of colour re-connections between strings.
Similar to the model described above one can try to describe the rise of \meanpt\ in p--Pb
collisions by a superposition of parton scatterings from an incoherent superposition of p--nucleon collisions \cite{Bzdak:2013lva, Abelev:2013bla} (see Fig. \ref{fig:meanpt} middle panel,
solid line). 
The observed rise of  \meanpt\ is significantly stronger than expected suggesting that also in this case effects from interactions between MPIs like colour re-connections are at work. 
The EPOS generator 
which includes hydrodynamic collective flow
can reproduce the p--Pb data, however, it fails to describe peripheral Pb--Pb collisions.
ALICE has also measured the multiplicity dependence of \meanpt\ for identified particles ($\pi$, K, p, $\Lambda$).
Here, a clear mass ordering $\meanpt_{\Lambda} > \meanpt_{\rm p} > \meanpt_{\rm K} > \meanpt_\pi$ is observed \cite{Abelev:2013haa}, which
is an indication for collective expansion with a common velocity field. The same kind of mass ordering is also 
qualitatively expected from colour re-connections \cite{Ortiz:2013yxa}.

In heavy-ion collisions, the increase of \meanpt\ and its mass ordering find
their natural explanation in the collective radial expansion of the system \cite{Heinz}. 
This scenario can be corroborated
in the blast-wave framework with a simultaneous fit to all particles for each multiplicity bin \cite{blastwave}. 
From the fit one obtains the mean radial velocity $\langle \beta \rangle $ and kinetic freeze-out temperature $T_{\rm kin}$.
It is found that the trend of  $T_{\rm kin}$-$\langle \beta \rangle $ as a function of multiplicity is very similar in p--Pb and Pb--Pb collisions
\cite{Abelev:2013haa}. The same trend,  albeit at a 30\% smaller $T_{\rm kin}$, can be also reproduced with 
pp collisions simulated by PYTHIA8 when colour reconnections are included.

\begin{figure}
\begin{minipage}{0.48\linewidth}
\centering
\includegraphics[scale=0.4]{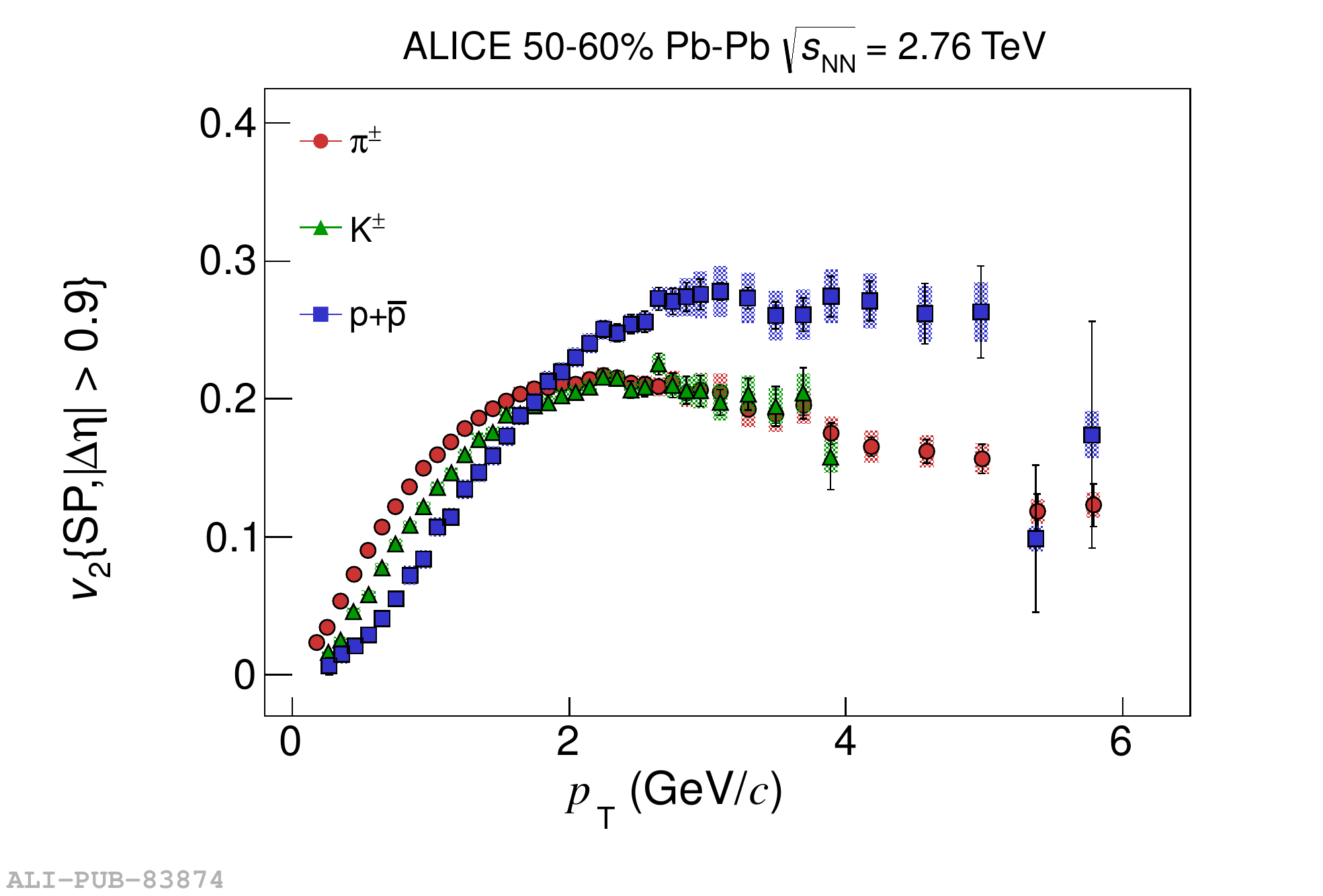}
\end{minipage}
\hspace{0.0 \linewidth}
\begin{minipage}{0.48\linewidth}
\centering
\includegraphics[scale=0.37]{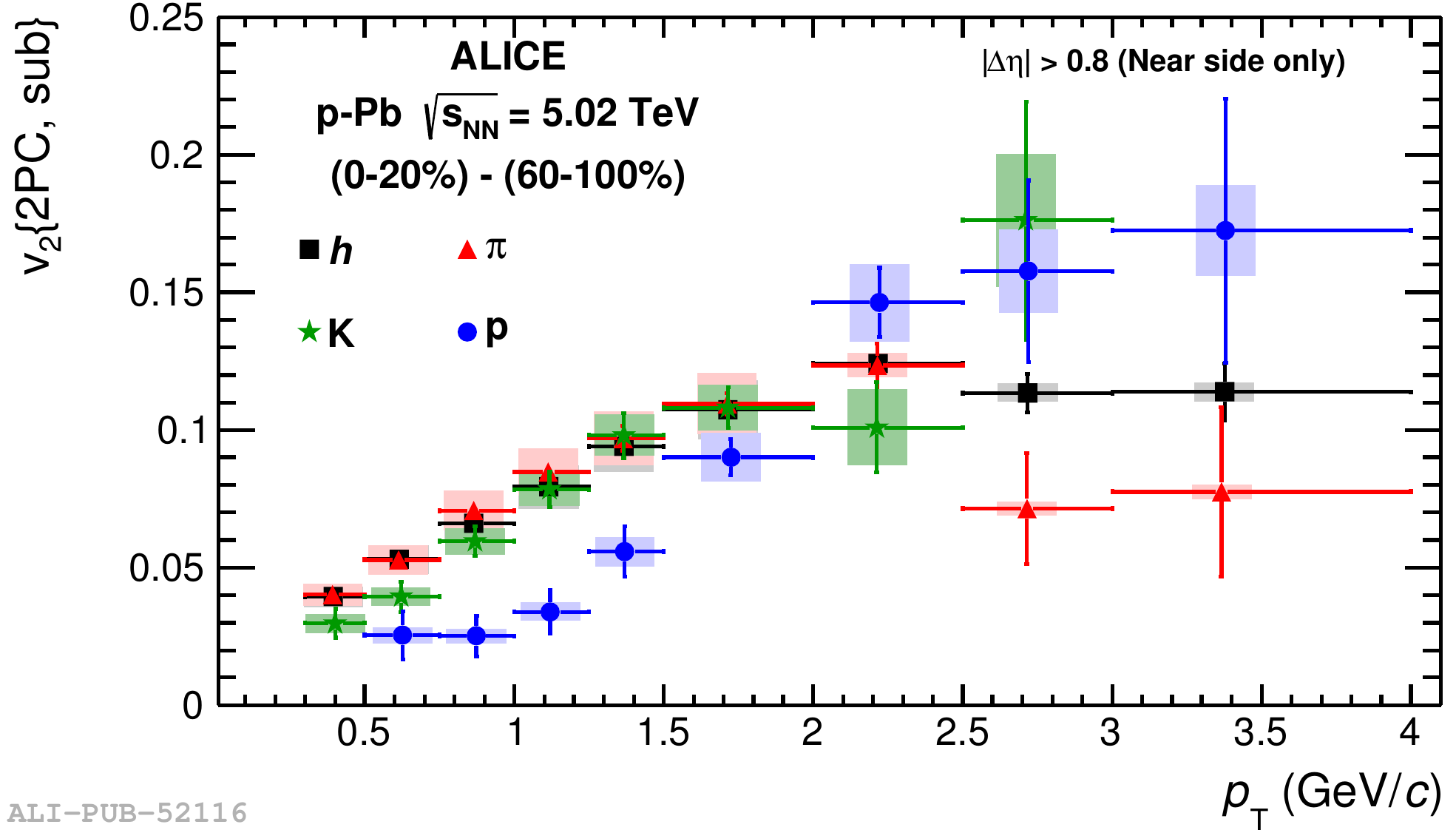}
\end{minipage}
\caption{For hadrons, pions, kaons and protons:
(left panel) $v_2$(\pT ) for mid-central Pb--Pb collisions \cite{Abelev:2014pua}
and (right panel) $v_2\{{\rm 2PC,sub}\}$(\pT ) from 2-particle 
correlations in the 0--20\% multiplicity class after subtraction of the
correlation from the 60--100\% multiplicity class \cite{ABELEV:2013wsa}. 
}
\label{fig:flow}
\end{figure}
\subsection*{Two-particle angular correlations in {\rm p--Pb}}
The strongest evidence for collective effects in p--Pb results from the study of triggered 2-particle angular
correlations in the azimuthal ($\Delta \varphi$) and pseudo-rapidity ($\Delta \eta$) differences.
These correlations show symmetric ridge like-structures elongated in $\Delta \eta$ at the near-side
($\Delta \varphi = 0$)  and away-side  ($\Delta \varphi = \pi$) of the trigger particle
\cite{CMS:2012qk, Abelev:2012cya, Aad:2012gla}.
They are very similar to the momentum anisotropy observed in Pb--Pb \cite{Abelev:2014pua}, 
where the effect is attributed 
to collectivity (flow) (Fig. \ref{fig:flow}) \cite{ABELEV:2013wsa}.
Whereas these correlations can be explained by  hydrodynamic flow or models based on the Color Glass Condensate 
\cite{Dusling:2012wy},
the above mentioned MPI Monte Carlo models based on colour re-connections are not yet able to describe this observation.  

\begin{figure}[htbp]
\centering
\includegraphics[scale=0.72]{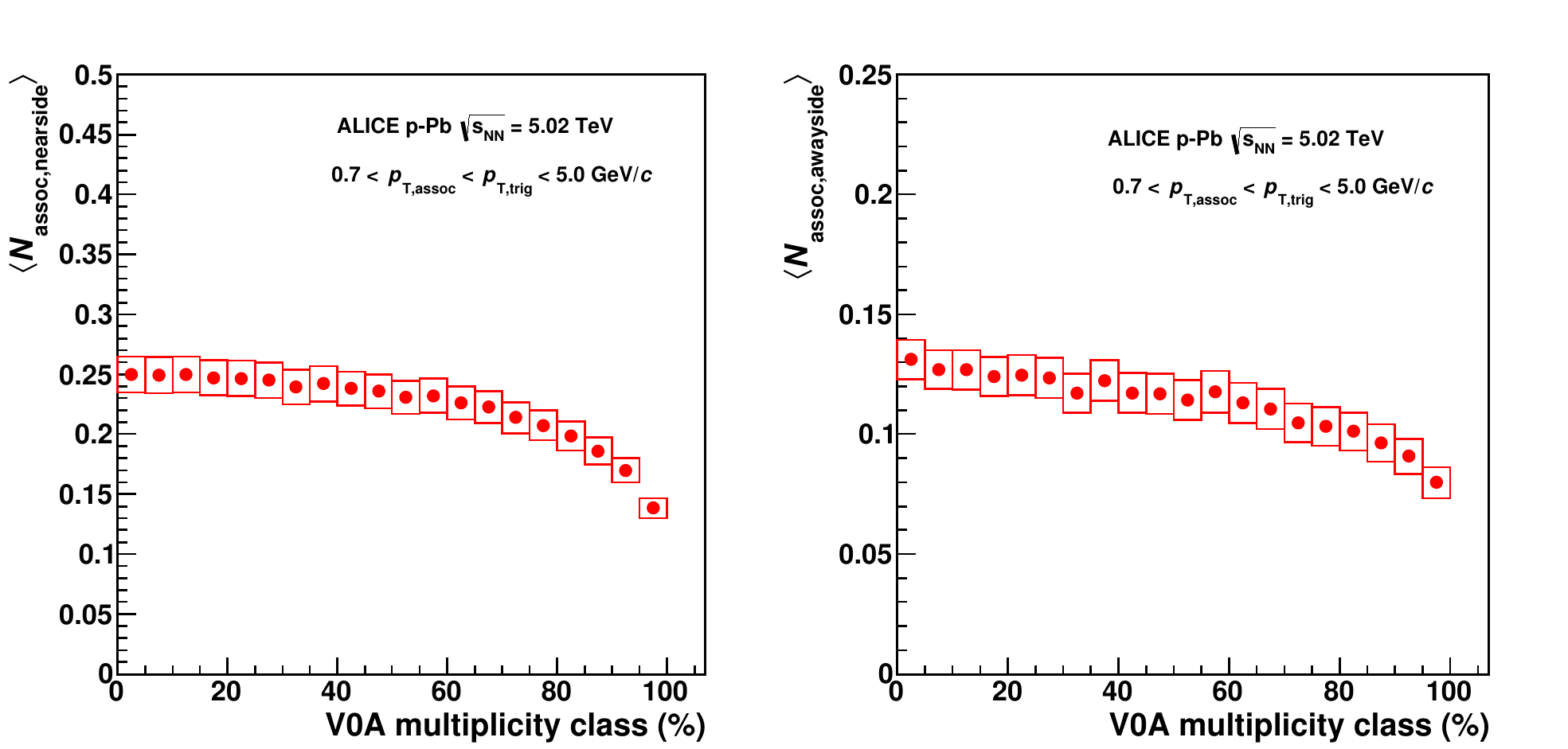}
\caption{
Near-side (left) and away-side (right) per-trigger yield as a function of V0A multiplicity class in p--Pb collisions 
at $\snn = 5.02$~ TeV with $0.7 < \pTa <  \pTt < 5.0$~GeV/$c$ \cite{Abelev:2014mva}.
} 
\label{fig:corrpa}
\end{figure}

Whereas $\meanpt $ and long-range correlations are sensitive to collective effects, possible local changes to parton fragmentation can be studied by jet-like short range correlations in 
$\Delta \eta - \Delta \varphi$. Figure \ref{fig:corrpa} shows the near-side and away-side per trigger yields
as a function of the V0A multiplicity class in p--Pb collisions at $\snn = 5.02$~TeV \cite{Abelev:2014mva}.
Contrary to the strong event activity dependence observed for \meanpt\  and for the long-range 
correlations, the yields measured in jet-like correlations stay unmodified over a wide range of V0A multiplicity percentile. Only for the lowest
multiplicities ($> 70\%$) a significant decrease of the  yields is observed.
It can be expected that this observation puts new  constraints on models implementing 
coherence effects between MPI or collective hadronization.

\subsection{Summary}
ALICE has explored the rich phenomenology of MPI in pp collisions. In particular, the 
study of the multiplicity dependence of jet-like azimuthal two-particle correlations shows in 
a quite direct way that,
in agreement with Monte Carlo simulations, high multiplicity pp events can be understood
as a superposition of semi-hard scatterings.

Studies of observables as a functions of multiplicity have been extended to p--Pb.
Phenomena  which are in 
nucleus--nucleus well established as the consequence of collective hydrodynamic flow are observed.
The observed increase of \meanpt\ and its mass ordering has been also observed in pp
where it can be attributed to colour re-connections between strings formed by multiple parton-parton 
interactions. It can be speculated that this effect plays also an important role in p--Pb where the MPIs
overlap in a transverse region similar to pp. 
The EPOS Monte Carlo generator 
which includes collective fragmentation effects via hydrodynamic flow for small collision systems
is in agreement with  the pp and p--Pb data. Hydrodynamic flow is also a natural explanation for the 
symmetric double-ridge structure observed in p--Pb.
With the aim to get more insight into the possible influence of these effects on parton fragmentation,
ALICE has studied 
 jet-like angular correlations at low \pT\  as a function of multiplicity in p--Pb.
Surprisingly the fragmentation properties expressed as the yield associated to a trigger particles stays 
constant over a wide range of multiplicities. Future model comparisons  will show whether these findings can be used to further constrain coherent fragmentation effects.



\end{document}